# A Double Machine Learning Trend Model for Citizen Science Data

Daniel Fink[1*], Alison Johnston[2], Matt Strimas-Mackey[1], Tom Auer[1], Wesley M. Hochachka[1], Shawn Ligocki[1], Lauren Oldham Jaromczyk[1], Orin Robinson[1], Chris Wood[1], Steve Kelling[1], and Amanda D. Rodewald[1]

[1]Cornell Lab of Ornithology, Cornell University, Ithaca, NY 14850, USA.
[2]Centre for Research into Ecological and Environmental Modelling, School of Maths and Statistics, University of St Andrews, St Andrews, UK.
*Corresponding author. daniel.fink@cornell.edu

**Abstract**
1. Citizen and community-science (CS) datasets are typically collected using flexible protocols. These protocols enable large volumes of data to be collected globally every year, however, the consequence is that these data typically lack the structure necessary to maintain consistent sampling across years. Thus, it is not possible to estimate accurate population trends with simple methods, because population changes over time are confounded with changes in the observation process.
2. Here we describe a novel modeling approach designed to estimate species population trends while controlling for the interannual confounding common in citizen science data. The approach is based on Double Machine Learning, a statistical framework that uses machine learning methods to estimate population change and the propensity scores used to adjust for confounding discovered in the data. Additionally, we present a simulation method to identify and adjust for residual confounding missed by the propensity scores. Machine learning makes it possible to use large feature sets to control for confounding and model heterogeneity in trends.
3. To illustrate the approach, we estimated species trends using data from the CS project eBird. We used a simulation study to assess the ability of the method to estimate spatially varying trends in the face of real-world confounding. Results showed that the trend estimates distinguished between spatially constant and spatially varying trends. There were low error rates on the estimated direction of population change (increasing/decreasing) at each location and high correlations on the estimated magnitude of population change.
4. The ability to estimate spatially explicit trends while accounting for confounding inherent in citizen science data has the potential to fill important information gaps, helping to estimate population trends for species and/or regions lacking rigorous monitoring data.

**SECTION 1: Introduction**
Information on population trends is essential for conservation monitoring and management. To date, the estimation of interannual trends has largely been restricted to the analysis of data from structured surveys where, ideally, the same observers follow the same survey protocols at the same locations, dates and times each year. This controlled survey structure is used to minimize the interannual variation in the observation process that can lead to confounding. This same structure enables estimation using regression models that do not explicitly account for confounding (e.g. Kery & Royle, 2020; Link et al., 2020). However, these survey requirements also make it difficult to collect species-observation data at the scales necessary to monitor large groups of species across broad spatial extents, and at arbitrary times of year.

Citizen science projects are collecting increasingly large volumes of data on a variety of taxa (Pocock et al., 2017); however, the opportunistic approach typically taken towards data collection makes these data susceptible to interannual changes in the observation process. For example, several studies have documented interannual variation in spatial site selection (August et al. 2020, Shirey et al. 2021, Zhang et al. 2021). Participant populations change as new participants join projects and continuing participants improve the way they conduct surveys (Johnston et al., 2022). Data collection protocols change, either through deliberate choice or for uncontrollable reasons. Examples of deliberate changes are those caused by the use of short-term incentives or games (Xue et al., 2016), and in the long-term promotion of 'best practice' protocols (e.g. submission of complete checklists; (Sullivan et al., 2009). Uncontrollable changes include improvements to equipment, such as binoculars, the development of species identification apps, and external forces shaping observers' behaviour, such as the COVID pandemic (e.g. Hochachka et al., 2021). Studies of citizen science data have also shown how interannual changes in the observation process bias trend estimates. Bowler et al. showed how changes in the spatial site selection produce species-specific biases in trend estimates. Zhang et al. 2021 showed how bias can arise despite interannual survey structure, documenting how unexpected changes in survey censoring due to urbanization biased trends in species richness. Thus, a key challenge for using citizen science data for trend estimation is controlling for the plethora of potentially confounding sources of interannual variation.

Recent analytical developments in other fields have created an opportunity to use citizen science data for trend estimation whilst controlling for interannual confounding. Double Machine Learning (DML) is a statistical framework developed to utilize generic Machine Learning (ML) methods (e.g., random forests, lasso, penalized-regression, boosted models, deep neural networks) for causal inference (Chernozhukov et al., 2018b). The DML framework has been increasingly used to estimate heterogenous treatment effects within large, feature-rich observational datasets, an important problem in many disciplines where confounding is a central concern, from economics (Athey, 2017) to personalized medicine (Obermeyer & Emanuel, 2016).

We consider DML for estimating spatially explicit species population trends from citizen science data. Conceptually, DML divides trend estimation into three separate modeling tasks. The goal of the first task is to predict local population sizes averaged across the study period. To do this, a species distribution model is trained to learn how observations of species vary with a set of features (e.g. climate, landcover, search effort). The goal of the second task is to identify confounding sources of variation in the data. To do this, a propensity score model is trained, which describes how the features vary systematically over time (Ramsey et al., 2019). In the third task, the expected population size and observation year are used as benchmark values to help isolate the trend so it can be estimated without the influence of confounding features. The inclusion of spatial features in the third prediction task creates the opportunity to generate spatially explicit trend estimates. The ability to capture trends with high resolution (e.g. landscape scale) is valuable for studying the processes affecting populations at these scales (e.g. agriculture, energy development, urbanization) (Rose et al., 2017).

Developing the propensity score model, plays a critical role in identifying the patterns of interannual variation in sampling that can lead to confounding bias. While the ability to include large feature sets in the propensity score model can provide broad, detailed control for confounding, it is nevertheless important to be able to quantify the degree to which this model adequately captures the interannual variation in sampling. However, we are not aware of any standard diagnostics for assessing the effectiveness of the propensity score model. We propose a novel simulation-based diagnostic tool to help identify residual confounding, i.e. confounding sources of variation in the available feature data missed by the propensity score model. Information gained from these diagnostic simulations can be used to adjust the trend estimates.

We use the DML trend model and residual confounding for a real-world application estimating population trends with data from eBird, a popular citizen science project that has been collecting bird observation data since 2002 (Sullivan et al., 2014). eBird engages large numbers of participants who each decide where, when, and how to participate. As with many other citizen science projects, the limited structure in eBird has given rise to an evolving, heterogenous observation process where interannual confounding is a central concern when estimating population trends. The goal is to estimate the average annual rate of change in breeding season abundance 2007–2021 at a 27km resolution across North America for three species of birds with different distributions and habitat preferences: wood thrush (*Hylocichla mustelina*), Canada warbler (*Cardellina canadensis*), and long-billed curlew (*Numenius americanus*).  Species-specific simulation studies were used to assess overall performance and the ability of the method to estimate spatially varying trends.

**SECTION 2: The DML Trend Model**
In this section, we introduce the DML trend model and our proposed simulation-based residual confounding analysis. Then we present a simple synthetic example to demonstrate how interannual confounding can be controlled using this approach.

**SECTION 2.1 Double Machine Learning**
To estimate population trends, we begin with the model that describes variation in species abundance $Y$, the response variable (also called the *outcome* or *label* variable), as
$$Y = \tau T + \mu(X) + \epsilon. \tag{1}$$
The objective is to estimate parameter $\tau$, the rate of change in abundance per unit time $T$. For convenience we assume $Y$ is a real-valued measure or index of species abundance, but integer counts, or binary indicators of a species' occurrence can be accommodated without loss of generality. We also assume that $T$ measures time in units of years and that $\tau$ is the interannual trend, but other units can be accommodated to estimate trends over different time scales without loss of generality. The function $\mu$ is a non-parametric function of the vector $X = (X_1, \dots, X_k)$ consisting of the features (also called *covariates* or *predictor* variables) that capture effects that are constant across years. Features in $X$ can include both ecological process variables (e.g. habitat or climatic conditions) and observation process variables (e.g. search effort or survey time of day). The number of features, $k$, can be large. The variable $\epsilon$ is a stochastic error term.

To understand how confounding can affect trend estimation in (1) it is useful to consider an idealized data set where the features $X$ include all important sources of variation in abundance and the observations are collected as a random sample across $X$, independently drawn each year of the study period. Under these conditions, $E[\epsilon|X, T] = 0$ and the trend $\tau$ can be estimated without bias. In practice, confounding can arise when there are systematic year-to-year changes

in the observation process that affect the distribution of $X$. For example, surveys could be conducted in sites with better habitat quality over time. In such situations, biased estimates of $\tau$ would result from employing estimation methods that do not account for the confounding, because all inter-annual changes are ascribed to $\tau$ in equation (1) (Imbens & Rubin, 2015).

A common strategy to account for confounding when analyzing non-experimental data uses propensity scores to adjust estimates (Rosenbaum & Rubin, 1983). In this approach a propensity score model is introduced to keep track of confounding, which can be framed here as the dependence of $T$ on features $X$. The propensity score model is written as,
$$T = s(X) + \delta, \qquad (2)$$
where $s$ is a non-parametric function of the features $X$ and $\delta$ is a stochastic error term where $E[\delta|X] = 0$.

DML solves equations (1) and (2) using the plug-in estimator of (Robinson, 1988). The plug-in estimator is constructed by substituting the conditional mean response averaged across T,
$$m(X) = E[Y|X] = \tau\, s(X) + \mu(X), \qquad (3)$$
into equation (1) to yield the residual-on-residual regression,
$$\bigl(Y - m(X)\bigr) = \tau\,\bigl(T - s(X)\bigr) + \epsilon. \qquad (4)$$
The residual on the left side of (4) isolates the change in abundance by regressing out year-invariant effects of features $X$ on $Y$ and the residual on the right side removes the effects of confounding by regressing out the effects of features $X$ on $T$. This formulation motivates the plug-in estimator where $m(X)$ and $s(X)$ are separately predicted and then plugged into equation 4 to estimate the trend, $\tau$.

Chernozhukov et al. (2018) showed how the plug-in estimator can accurately estimate $\tau$ even when the predictions of $m(X)$ and $s(X)$ are noisy and suffer from regularization bias. This makes it possible to take advantage of large feature sets $X$ using generic statistical and machine learning methods (e.g. penalized regressions, lasso, random forests, boosted models, deep neural networks and ensembles of these methods). The ability to include large, rich feature sets is practically important because it can improve inference by strengthening the conditional mean model while providing for strong confounding control through the propensity score model.

In this paper we used Causal Forests (Athey et al., 2019), an implementation of the DML framework that uses Random Forests (Breiman, 2001) as the machine learning model for each of the prediction tasks. Moreover, by considering the trend $\tau$ to be a non-parametric function of the feature vector $W = (W_1, \ldots, W_m)$, where the number of features $m$ can be large, Causal Forests extend the response model to
$$Y = \tau(W)T + \mu(X) + \epsilon. \qquad (5)$$
The ability to estimate trends conditional on a set of features $W$ can be used to identify and study heterogeneity in population change. For example, by including spatial features in $W$, spatially explicit trends can be estimated. In the causal inference literature, equation 5 is known as a *heterogenous treatment effect* or *conditional average treatment effect* estimator. In statistics it is equivalent to a varying coefficient model for $\tau$ (Hastie & Tibshirani, 1993). (Please see SI Section S1 for a brief review of DML literature)

The goal in this paper is to use DML to strengthen inference about population trends by controlling for interannual confounding, but with additional assumptions it can also be used for causal inference. The models in equations 1 and 5 are closely connected to the potential outcomes framework (Rubin, 1974) that describe the conditions necessary for casual inference. Within this framework, DML and Causal Forest estimators are unbiased and normally distributed (Athey et al., 2019; Chernozhukov et al., 2018a). Thus, unlike many machine learning methods, DML can be used for statistical and causal inference. (See SI Section S2 for more information about the potential outcomes framework.)

**SECTION 2.2 Residual Confounding**
Propensity scores provide a theoretically justified strategy to control for confounding in $X$ but in practice good performance requires that the model $s(X)$ accurately describes the confounding. Model and data limitations can make this challenging. When $s(X)$ fails to fully capture the confounding then there will be what we call *residual confounding* with respect to $X$. Residual confounding can arise if $X$ are well suited to explain patterns of year-invariant variation in abundance but are not well suited to reveal interannual variation. For example, search duration, the amount of time spent searching for species, is often an important predictor of abundance. If the amount of intra-annual variation in search duration is large compared to the inter-annual changes, then the latter may be more difficult to predict. However, we are not aware of any standard diagnostics for assessing the effectiveness of the propensity score model.

Here we propose the use of a simulation-based diagnostic for residual confounding in the DML model. First, we simulate datasets $\{X^*, W^*, T^*, Y^*\}$ with a known species population trend $\tau^{sim}$ while maintaining all the interannual confounding in the original features. Then to diagnose residual confounding, we look for systematic differences between $\tau^{sim}$ and $\tau^*$, the DML trend estimate based on the simulated data. The differences $(\tau^{sim} - \tau^*)$ can be used to adjust the estimate.

Simulating data from a null model, i.e., zero-trend, while maintaining all the interannual confounding in the original features is straightforward and such data can be computed for any DML estimate in two steps:
1. Generate a synthetic feature set $\{X^*, W^*, T^*\}$ by resampling with replacement from $\{X, W, T\}$ stratified by $T$, and, then,
2. Compute synthetic responses $Y^* = m(X^*)$ based on the conditional mean model (3).

The first step is to generates realistic feature sets while maintaining the interannual variation in $X$. Using the conditional mean model (3) in the second step ensures that the synthetic data have zero trend while maintaining year-invariant patterns of variation in abundance associated with $X$. Moreover, because these synthetic data are based on a zero or null-trend model, additional assumptions (beyond those of the DML) about the unknown trend are avoided.

We illustrate the residual confounding diagnostic and adjustment in the synthetic example presented in the next Section. In Section 3.4 we extend the use of residual confounding simulation to also assess DML trend estimates across a suite of simulated non-zero trends.

**SECTION 2.3 Synthetic Example**
We present a simple example to illustrate how the DML trend model uses propensity scores and the residual confounding adjustment to control for interannual confounding. To do this, we generated fully synthetic data sets $\{X, W, T, Y\}$ with known structure and then compared DML trend estimates based on the synthetic data sets with and without adjustments.

To generate the synthetic data, the log abundance $Y_i$ reported on the $i^{th}$ survey was specified using the following response model,

$$Y_i = \alpha_0 + \sum_{j=1}^{J} \alpha_j h_{ji} + \sum_{j=1}^{J} \beta_j o_{ji} + T_i \left( \tau_0 + \sum_{j=1}^{J} \tau_j w_{ji} \right) + \epsilon_i, \qquad (6)$$

where $h_{ji}$ was the $j^{th}$ habitat feature, $o_{ji}$ was the $j^{th}$ observation process feature, $w_{ij}$ was the $j^{th}$ trend feature, $T_i$ was the year of the $i^{th}$ survey, and $j = 1, \ldots, J$. Additionally, the synthetic data were constructed with two confounding sources of variation,

$$\tilde{T}_i = h_{1i} + o_{1i} + \delta_i, \tag{7}$$

where participants 1) selected sites with more habitat type $h_1$, and 2) spent increasing amounts of effort $o_1$ searching for species at later times $T$. The continuous time variable $\tilde{T}$ was transformed into discrete years $T = 0, \ldots, 10$ using empirical quantiles. Terms $\epsilon_i$ and $\delta_i$ were independent stochastic error terms.)

To mimic problems where the availability of large, potentially informative feature sets motivates the use of machine learning, we set $J = 25$ but with only three non-zero parameters, ($\alpha_1 = 10, \beta_1 = 10, \tau_0 = -0.05$). These parameters were chosen to generate data sets with moderate confounding and where the inter-annual variation in the species population size $\tau_0$ was relatively small compared to the intra-annual variation (Fig. 1A).

These are common challenges in many real applications. For example, the spatial variation in species abundance across its range may be greater than the interannual variation of the trend itself, especially for trends over short time periods. Similarly, the heterogeneity of the citizen science observation processes may generate high levels of confounding variation. Here, the interannual increases in survey coverage and search effort (equation 7) both generate positive biases because they are associated with reporting of higher species abundance (Equation 6). The difference in slopes between $\tau$ and the trend fit with a simple linear regression show the effects of these biases (Fig 1A).

We generated 100 synthetic datasets and estimated trends using Causal Forests with the package *grf* (Tibshirani, et al. 2021). Each of the synthetic datasets had a sample size of 1100 consisting of 100 surveys per year. The 50 features $X = \{h, o\}$ were used to train the random forests used to predict the conditional means and propensity scores. The 25 trend features $W$ were used to train the random forest trend model.

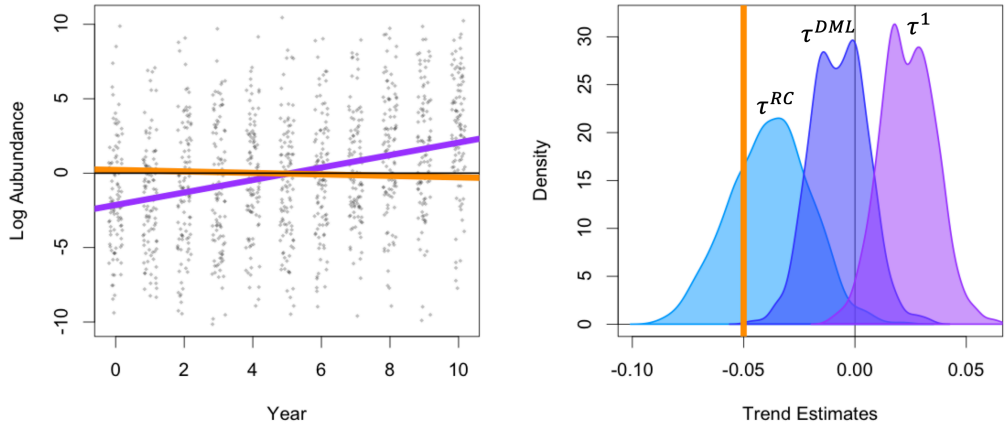

*Figure 1: Synthetic example. (A) Scatterplot of a single realization of simulated log abundances for years 0-10 with site selection and search effort confounding. The simulated trend with slope $-0.05$ (orange line) reflects a small portion of the overall variation in abundance. The difference in slopes between the simulated trend and the trend fit with a simple linear regression (purple line) reflects the effects of confounding. (B) Density plots across 100 realizations for trend estimates $\tau^1$ without any correction for confounding (purple), trend estimates $\tau^{DML}$ with the propensity score adjustment (dark blue), trend estimates $\tau^{RC}$ with the propensity score and the*

*residual confounding adjustment (light blue) with the (orange) vertical line showing the simulated trend at $-0.05$.*

To assess the impact of the propensity score and residual confounding adjustments we computed four trend estimates for each synthetic data set:
1) $\tau^1$: The Causal Forest trend *without* adjusting for the propensity scores. This was computed by using $mean(T)$ as a constant value for the propensity scores,
2) $\tau^{DML}$: The Causal Forests trend *with* propensity score adjustment,
3) $\tau^*$: The Causal Forests trend with propensity score adjustment based on data simulated from the conditional mean null model (Section 2.2),
4) $\tau^{RC} = \tau^{DML} - \tau^*$: The Causal Forests trend with propensity score and residual confounding adjustments.

Without any adjustment for confounding, trend estimates $\tau^1$ were strongly biased (Fig. 1B, purple distribution). Including the propensity score adjustment reduced but did not completely remove the confounding bias (Fig. 1B, dark blue distribution). The residual confounding adjustment provided additional control for confounding (light blue distribution). Together, these adjustments have debiased $\tau^{RC}$ enough for the 95% confidence interval to cover the true trend in 94% of the simulations.

The differences between $\tau^{DML}$ and $\tau^{RC}$ reflect the difficulty of modeling the propensity scores and conditional means using Random Forests to identify sparse, noisy, linear signals with only moderate sample sizes. In general, the residual confounding adjustments function to ensure control of confounding when the propensity score model suffers from deficiencies. The residual confounding adjustments will be strongest when the propensity score model is weak and will tend to zero the stronger the propensity scores model is. Thus, we expect residual confounding will be most valuable when the set of features is limited, signal to noise ratios are low, or when confounding is not well understood.

**SECTION 3: Trend Analysis of North American Breeding Birds**
In addition to the synthetic example in Section 2.3 we also estimate trends based on data from eBird, a popular citizen science project that has been collecting bird observation data since 2002 (Sullivan et al., 2014). This application presents the challenges of estimating spatially explicit trends in abundance across large geographic extents in the face of confounding and temporally correlated observations.

In this section we describe the eBird data, the species abundance model underlying trend estimation, the Causal Forest implementation, the residual confounding analysis, and the species-specific simulation study used assess the performance of the method. All computing was done in the R statistical computing language (R Core Team, 2019).

**SECTION 3.1: Data**
eBird is a semi-structured survey (Kelling et al., 2019) because its flexibility allows participants to collect data in the ways they choose, but ancillary data are collected that describe the variation in data collection methods. To help control for variation in observation process, we analyzed the subset of the data where participants report all bird species detected and identified during the survey period, resulting in *complete checklists* of bird species. This limits variation in preferential reporting rates across species and provides a basis to infer species non-detections. We also required all checklists to include key ancillary variables describing characteristics of each birdwatching event, for example the time of day and distance travelled. These variables and others can be used to adjust for variation in detection rates (Johnston et al., 2019).

We calculated trends for three species that represent a range of different breeding niches, observation processes, and processes driving population change. Wood thrush is a commonly reported bird of the deciduous forest in eastern North America. Canada warblers are a less commonly reported forest bird that breed in the boreal forests of North America. Long-billed curlew is an infrequently reported shorebird that breeds in the grasslands of the arid interior of North America.

We analyzed eBird data from 2007 to 2021 within each species' previously identified breeding range and season (Fink, Auer, Johnston, Strimas-Mackey, et al., 2020). To prepare the data for the trend analysis we aggregated data using a (27km × 27km x 1week) grid based on checklist latitudes, longitudes, and dates. We computed grid cell averages for four classes of information: (1) The number of individuals of the given species reported in each grid cell was used as the response variable ($Y$); (2) Five observation-effort features describing how participants conducted surveys were used as features to account for variation in detection rates (Johnston et al., 2019); (3) 14 features describing short-term temporal variation — date, time of day, and hourly weather — were used as features to account for variation in availability for detection; (4) A suite of 57 spatial features describing the composition and configuration of landscapes in each grid cell were used to capture associations between species and elevation, topography, land & water cover, land use, hydrology, and road density. Please see the SI Section S3 for details about data and data processing.

**SECTION 3.2 Species Abundance Model**
Species expected abundance can be defined as the product of the species' occurrence rate and the expected count of the species given occurrence, within a given area and time window (Zuur et al., 2009). Based on this definition and the chain rule, the rate of change in species abundance is the sum of two terms: 1) the rate of change in the species occurrence, and 2) the rate of change in species count given occurrence. Intuitively, trends in species abundance can arise from trends in the occurrence rate (e.g., as a function of whether the habitat is even suitable for a species) and/or trends in expected counts given occurrence.

We estimated each trend component with its own DML. To estimate the interannual rate of change in occurrence rate a Causal Forest was trained based on the binary response variable indicating the detection/non-detection of the species and the features. Then to estimate the interannual rate of change in the log transformed species count, a separate Causal Forest was trained based on the continuous response variable (log transformed count) and features, using the subset of surveys where the species was detected (i.e., all counts were positive).

To quantify the sampling variation in abundance trend estimates arising jointly from the estimated trend in species occurrence rates and the estimated trend in species counts, given occurrence, we adopted a data resampling approach and computed an ensemble of 100 estimates. We calculated 80% confidence intervals using the lower 10th and upper 90th percentiles across the ensemble. Additionally, averaging estimates across the ensemble provides a simple way to control for overfitting (Efron, 2014). Please see the SI Section S4 for additional information about constructing the ensemble.

**SECTION 3.3 Causal Forest Implementation**
Causal Forests were fit using the *grf* package (Tibshirani et al., 2020) and were grown with 2000 trees using automatic parameter tuning for all parameters. The feature sets for the conditional mean and propensity score models included (1) observation effort, (2) short-term temporal, and (3) spatial features. We also included latitude and longitude as features in the marginal model to account for residual spatial patterns of abundance. To account for spatial variation in trends we included all spatial features in W along with latitude and longitude to account for residual patterns.

**Section 3.4: Residual confounding**
The simulation-based residual confounding assessment presented in Sections 2.2 and 2.3 was based on a simulated null model. For the eBird analysis we extended the simulations to also assess confounding bias under a range of important nonzero trend scenarios. (See Section 3.5 for a description of the scenarios.) We implemented the residual confounding diagnostics and adjustments at the species level because interannual variation in how participants conduct surveys can generate distinct biases for each species. The goal was to find a set of parameters to describe and correct for residual confounding that would generalize well for all locations in the species range, regardless of the direction, magnitude, or spatial pattern of the unknown trend. To do this we fit a linear regression model for each species, estimated using all locations within the species' range for all simulation scenarios. Predictions from this regression model were used to make residual confounding adjustments (See Figure 2).

**SECTION 3.5: Simulation Study**
The species-specific simulations were constructed to create data meeting four objectives for each species: 1) realistic patterns of year-invariant patterns of occurrence and counts on eBird checklists; 2) with specified trends in abundance; 3) including temporal correlation that arises from environmental stochasticity in population growth rates; while 4) maintaining the interannual confounding in the original eBird data.

To assess the overall performance in detecting and describing spatial trend patterns, trends were simulated at a 27km × 27km spatial scale across each species range and across 10 scenarios with zero and non-zero trends varying in direction, magnitude, and spatial pattern. Magnitudes were set to <1% (weak), 3.3% (moderate) and 6.7% per year (strong) based on IUCN Red List criteria (IUCN, 2019). The spatially varying trends were constructed to vary in direction and magnitude along a gradient from the core to the edge of the species' population (Figures 3, SI-3, and SI-4). All scenarios also included temporally correlated stochasticity as expected from interannual environmental factors, an important characteristic of species abundance data.

The simulations were used for two tasks: 1) The training task to compute the residual confounding estimates, and 2) The testing task to evaluate the trend estimates after accounting for residual confounding. To maintain independence among these tasks, we independently generated two sets of simulated data sets for training and testing. Each simulated training data sets contained data from one of five different simulation scenarios (one null and 4 with varying magnitude, spatially constant and variable trends). Ten datasets were independently generated for each of the 5 scenarios. Thus, the simulated training data included 50 data sets generated under 5 different trend scenarios. Fifty simulated test data were generated independently of the training data, from 5 comparable but different scenarios. See the SI S5 for details about the simulations.

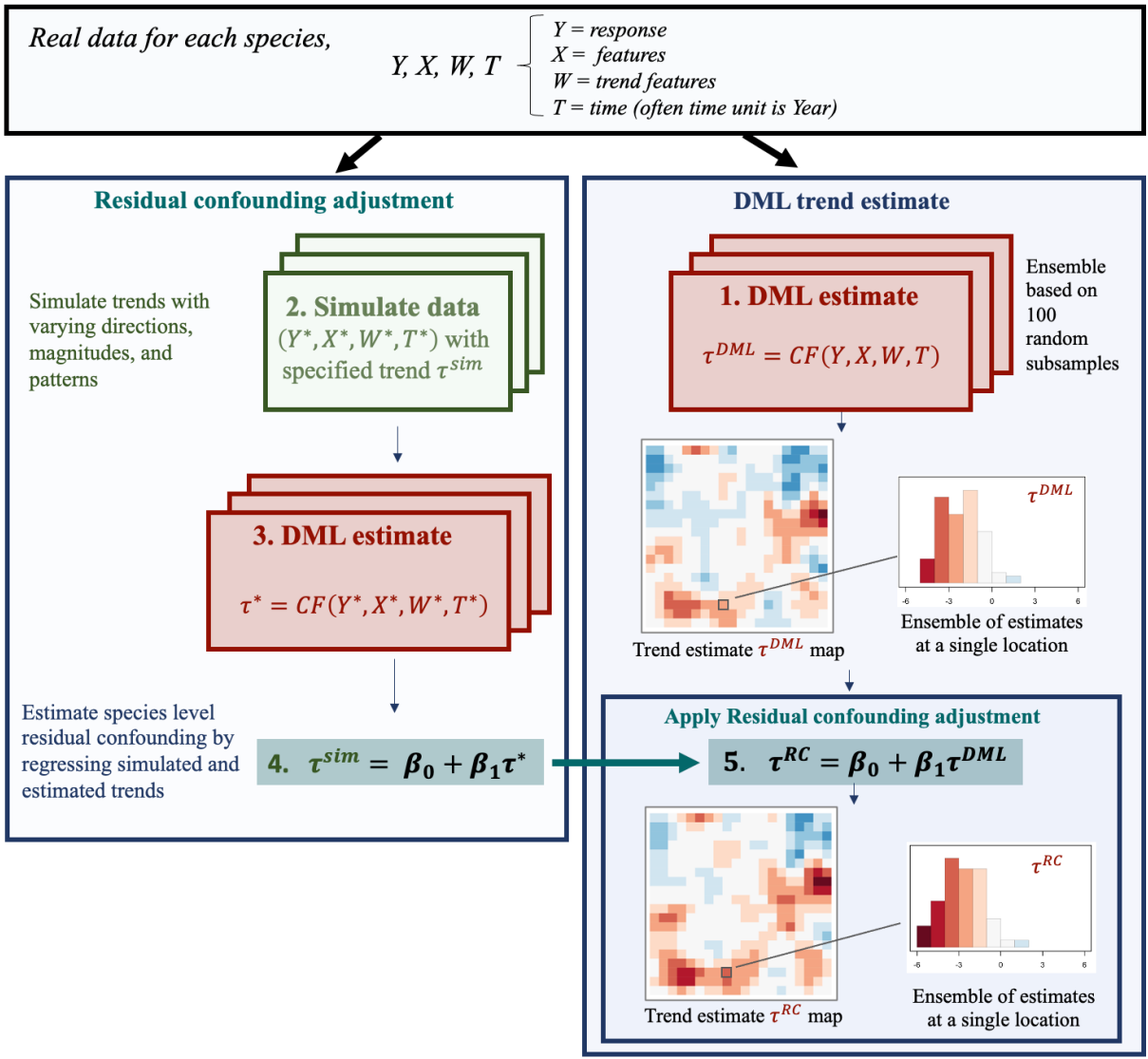

*Figure 2: A schematic workflow for the DML abundance trend model with residual confounding adjustment.* This schematic is based on the eBird analysis with numbers corresponding to the steps in Section 3.5. The residual confounding adjustment (LEFT) is based on the systematic differences between simulated trends (Step 2; expanded in Figure SI-1) and the DML estimates of these trends (Step 3, red boxes). Step 4 estimates the residual confounding coefficients by formally linking Step 2 and Step 3. The real data analysis (RIGHT) begins by calculating the DML trend estimate (Step 1) and then adjusts for residual confounding (Step 5). An ensemble of trend estimates is generated at each location, from which the point estimates and confidence intervals are calculated.

**Analysis workflow**
To compute each species' DML trend estimate with the residual confounding adjustment we followed the steps in the schematic workflow shown in Figure 2:
1) Estimate $\tau^{DML}$ using Causal Forests with propensity score adjustments based on the original data $\{X, W, T, Y\}$ (Sections 3.2 and 3.3),
2) Simulate data $\{X^*, W^*, T^*, Y^*\}$ with specified trends $\tau^{\text{sim}}$ (Section 3.5),
3) Estimate $\tau^*$ using Causal Forests with propensity score adjustment based on the simulated training data $\{X^*, W^*, T^*, Y^*\}$,
4) Fit $\tau^{\text{sim}} = \beta_0 + \beta_1 \tau^*$, the residual confounding regression (Section 3.4),
5) Apply the residual confounding adjustment to the original DML estimate: $\tau^{RC} = \beta_0 + \beta_1 \tau^{DML}$ using simulation-based parameters $(\beta_0, \beta_1)$ to adjust $\tau^{DML}$ based on the original data.

**Section 3.5.1: Confounding bias in eBird**
To measure the strength of the confounding bias and the performance of the propensity score and residual confounding adjustments we performed a simulation analysis comparing the trend estimates, $\tau^1$, the Causal Forest estimate *without* adjusting for the propensity scores to $\tau^{DML}$ and $\tau^{RC}$, both computed using the workflow above.

To measure the performance of the trend estimates we used the simulated test data with the specified trends $\tau^{\text{sim}}_{\text{test}}$ as the original data. This allowed us to assess the species-level confounding bias by fitting the regression $\tau^{\text{sim}}_{\text{test}} = \alpha_0 + \alpha_1 \tau$ separately for each of the estimates $\tau = (\tau^1, \tau^{DML}$ and $\tau^{RC})$. The intercept parameter $\alpha_0$ measured the distance between a zero-trend estimate and the corresponding expected value of the simulated trend. Thus, the intercept described the bias when estimating the trend direction (increasing or decreasing), with a value of zero indicating no *directional bias*. The slope parameter $\alpha_1$ measured how simulated trends scaled with the direction and magnitude of the estimated trends, with a value of 1 indicating no *scaling bias*.

| Species | Estimator | PS | RC | Intercept $\alpha_0$ | Intercept SE | Slope $\alpha_1$ | Slope SE |
|---|---|---|---|---|---|---|---|
| **Wood thrush** | $\tau^{RC}$ | Yes | Yes | 0.295 | 0.005 | 1.045 | 0.001 |
| | $\tau^{DML}$ | Yes | No | 0.254 | 0.005 | 1.342 | 0.001 |
| | $\tau^1$ | No | No | -0.610 | 0.005 | 1.480 | 0.001 |
| **Canada warbler** | $\tau^{RC}$ | Yes | Yes | 0.641 | 0.011 | 0.931 | 0.002 |
| | $\tau^{DML}$ | Yes | No | -0.212 | 0.010 | 1.159 | 0.002 |
| | $\tau^1$ | No | No | -1.589 | 0.011 | 1.171 | 0.002 |
| **Long-billed curlew** | $\tau^{RC}$ | Yes | Yes | 0.119 | 0.011 | 0.998 | 0.002 |
| | $\tau^{DML}$ | Yes | No | -3.663 | 0.010 | 1.153 | 0.002 |
| | $\tau^1$ | No | No | -4.767 | 0.012 | 1.143 | 0.002 |

*Table 1: Species-level estimates of confounding bias. Slope and intercept estimates and standard errors (SE) are presented for each species for trend estimates $\tau^1$ without any correction for confounding, trend estimates $\tau^{DML}$ with the Propensity Score (PS) adjustment, and trend estimates $\tau^{RC}$ with the propensity score and the Residual Confounding (RC) adjustment.*

The impact of the propensity score on directional bias is assessed by comparing bias coefficients between $\tau^1$ and $\tau^{DML}$, where an improvement is $\alpha_0$ moving towards 0 and $\alpha_1$ moving towards 1. Estimates show that the propensity score adjustments reduced, though did not eliminate, directional bias for all three species (Table 1). The propensity score adjustment also reduced the directional bias for Wood thrush, with a smaller reduction for Canada warbler and a small increase for long-billed curlew (Table 1) (Table 1).

The impact of the residual confounding adjustment on directional bias is assessed by comparing bias coefficients between $\tau^{DML}$ and $\tau^{RC}$. The residual confounding adjustment strongly reduced both directional and scaling bias for long billed curlew but had smaller effects on the other two species (Table 1). For Canada Warbler the residual confounding adjustment increased the magnitude of the directional bias and slightly decreased the magnitude of the scaling bias. For wood thrush the residual confounding adjustment led to a slight increase in the magnitude of the directional bias and it decreased the magnitude of the scaling bias.

### SECTION 3.5.2: Estimate performance

Next, we assessed the performance of $\tau^{RC}$, the DML trend estimates computed with both the propensity score and the residual confounding adjustments, using the simulated test data as the original data in the workflow and then comparing estimates with the specified trends $\tau_{test}^{sim}$. We evaluated the quality of the estimated trend magnitude (the average percent-per-year (PPY) rate of change in abundance 2007-21) and the trend direction (increasing/decreasing), two important inferential objectives for population monitoring. Directional errors were defined to occur when trends were estimated to be significantly different from zero but were in the opposite direction to the simulated trend. We considered estimates to be non-zero if the 80% confidence interval did not contain zero. Because directional errors varied strongly with trend magnitude (Figure SI-1), we reported the mean directional error rate, binned into categories of trend magnitude (see Supplemental Information for more details about the directional error). We also computed Pearson's correlation between simulated and estimated trends for non-zero trend estimates. Finally, we assessed the coverage of the resampling-based uncertainty estimates as the percentage of all 27km locations where the estimated intervals contained the simulated trend value. All assessments were based on independent test set data.

The mean directional error rate among non-zero trends was low for all species (Table 2). The correlations among non-zero estimates and simulated true values $\tau_{test}^{sim}$ was high. As expected, the directional error rates increased and the correlations decreased with the volume of species' data that were non-zero counts; from wood thrush (a commonly reported species in a region with high data density), to Canada warbler (less commonly reported in regions with lower data density), to long-billed curlew (infrequently reported compared to Wood Thrush and Canada Warbler within a relatively low-data density region of the continent). Interval coverage increased with decreasing amounts of species data (note, both sample sizes and detection rates decrease among species), though it was markedly less than the nominal confidence 80% level for all species.

| Species | Trend Scenarios | Directional Error | Correlation | CI Coverage |
|---|---|---|---|---|
| **Wood Thrush** | All | 2.4% | 96% | 47% |
| | Constant | 1.9% | 97% | 44% |
| | Varying | 1.9% | 92% | 48% |
| **Canada Warbler** | All | 6.8% | 88% | 49% |
| | Constant | 5.1% | 92% | 45% |
| | Varying | 5.5% | 79% | 51% |
| **Long-billed Curlew** | All | 3.4% | 88% | 61% |
| | Constant | 2.5% | 91% | 56% |
| | Varying | 2.6% | 86% | 62% |

*Table 2: Trend estimate performance. The directional error, correlation, and interval coverage are presented for each species, averaged across all evaluation scenarios (All) the spatially constant evaluation scenarios (Constant), across the spatially varying scenarios (Varying).*

**SECTION 3.5.3: Identifying spatial trends**

Finally, we assessed model performance identifying spatial heterogeneity in trends. We compared model performance between spatially constant and spatially varying scenarios for each species (Table 2). The similarity in performance between constant and varying trends highlights the ability of the model to adapt to heterogenous trends. Figure 4 shows trend maps for Wood Thrush for a single realization of each simulation scenario (See Supplemental Information Figures SI3 & SI4 for Canada warbler and long-billed curlew). These maps show how the model adapted to simulations with different directions, magnitudes, and spatial patterns.

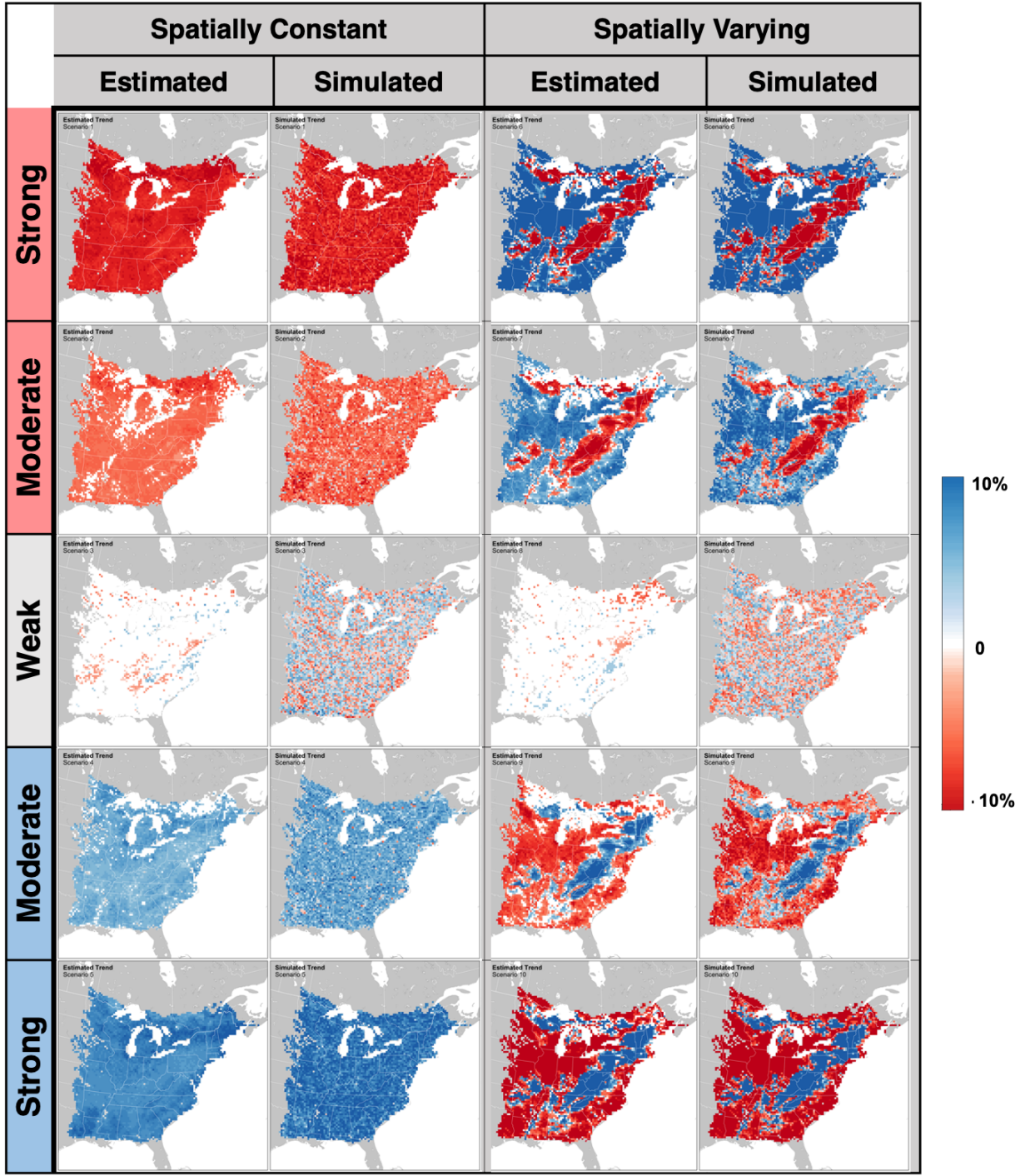

*Figure 3 Wood Thrush Trend Simulations.* All trend maps show the average annual percent-per-year change in abundance from 2007–2021 within 27km pixels (red=decline, blue=increase, white=80% confidence interval contained zero), intensity (darker colors indicate stronger trends). Simulated trends show scenarios varying by direction and magnitude along rows: weak (includes trends $\sim|1\%/yr|$), moderate (includes regions with trends $\sim|3.5\%/yr|$), and strong trends (includes regions with trends $\sim|6.7\%/yr|$). The columns show simulated and estimated trends for spatially constant and varying simulation scenarios.

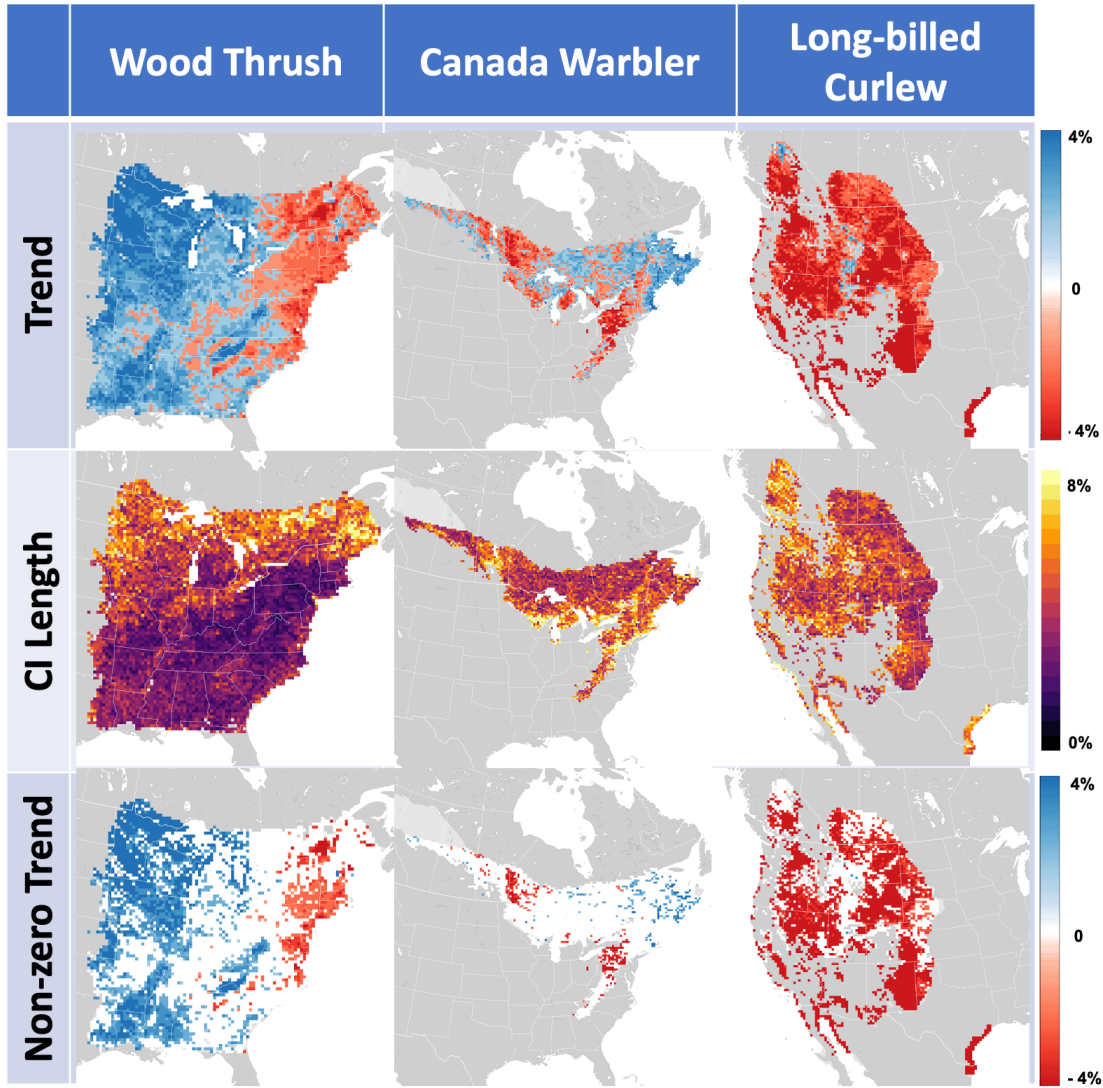

*Figure 4 Trend estimates Wood Thrush, Canada Warbler and Long-billed Curlew. All trend maps show the average annual percent-per-year change in abundance from 2007–2021 within 27km pixels (red=decline, blue=increase), intensity (darker colors indicate stronger trends). The top row shows the estimated trends, middle row shows confidence interval length, and the bottom panel shows the non-zero trends in red and blue with white in locations where 80% confidence interval contained zero.*

**SECTION 3.6: Species trend estimates**
Figure 4 shows maps of the estimated average annual percent-per-year change in abundance from 2007–2021 for all three species based on the real data. The Wood Thrush population shows steep declines in the northeast and increases in the southwest of its breeding season population, a pattern similar to other published studies (e.g. Fink, Auer, Johnston, Ruiz-Gutierrez, et al., 2020) The estimated population change for Canada Warbler also shows spatial patterning, though the uncertainty is relatively high. Long-billed Curlew shows strong, significant range-wide declines consistent with previous analysis (Rosenberg et al., 2019).

## SECTION 4: DISCUSSSION

Our work shows how Double Machine Learning can be used to estimate the interannual rate of population change from citizen science data that contain confounding variation through time. Simulation results showed that the propensity score adjustment reduced, though did not eliminate, confounding bias in eBird data. The simulation-based residual confounding adjustment provided further bias reductions. The resulting trend estimates accurately estimated trend direction and magnitude in most cases. These estimates were sufficiently accurate to distinguish between spatially constant and spatially varying patterns at a 27km×27km resolution, across multiple simulation scenarios.

Our study also highlighted several challenges with the DLM trend model including the task of modeling propensity scores, estimating uncertainty, and accounting for temporal and spatial correlation. In this section we discuss these challenges and how DML trend models may be used for other citizen science datasets and applications.

### SECTION 4.1: Confounding in Citizen Science data

We found that confounding bias is a challenge when estimating interannual trends with data from the citizen science project eBird. Without any correction for confounding, the simulation results showed that bias can be strong (4.7% directional bias for long-billed curlew), though it varied among species (0.6% directional bias for wood thrush). The ablation study showed that the propensity score adjustment performed by DML reduced bias, though did not completely remove it. The simulation results also demonstrated that the residual confounding adjustment made further reductions in bias.

### SECTION 4.1.1: Propensity Scores

These results highlight the important role of the propensity scores in the DML model. Adjustments based on propensity scores have the potential to provide strong bias control, but in practice, model and data challenges can limit their effectiveness. Even with eBird, a relatively well studied citizen science data set (e.g. Johnston et al., 2019), the residual confounding detected gaps in control. This highlights the importance of future work to improve our understanding of citizen science observation processes and how they evolve over time. Improvements in models and data will be valuable to improve the application of the DML trend model.

The goal of the propensity score model is to capture sources of interannual variation in the observation process that also impact the reported abundance of the species. This suggests a strategy for selecting features to include in the propensity score model where all important sources of variation in year-invariant abundance are included as the set of potential confounders. This is the strategy implicit our presentation where the features used for the conditional mean model are also used in the propensity score model. The propensity score model is not limited to this set of features, and we expect there are situations where it will be advantageous to consider others. However, we caution against indiscriminately including features in the propensity score models to avoid introducing or amplifying bias (Hernán & Robins, 2020). Again, this highlights the importance of future work to improve our understanding of citizen science observation processes and how they evolve over time.

### SECTION 4.1.2: Residual Confounding

In general, residual confounding adjustments function to ensure control of confounding when the propensity score model suffers from deficiencies. The residual confounding adjustments will be strongest when the propensity score model is weak and will tend to zero the stronger the propensity score model is. Thus, we expect residual confounding will be most valuable when the set of features is limited, signal to noise ratios are low, or when confounding is not well understood.

Critical to the success of any residual confounding analysis is the construction of the underlying simulated data. In general, the simulated data needs to have a known population trend while maintaining all the interannual confounding in the original features. The zero-trend, or null trend simulation in Section 2.2 is a convenient, general-purpose residual confounding analysis based on the conditional mean model. Thus, the applicability and performance of this approach will depend strongly on the quality of the conditional mean model.

For the eBird data analysis, it was also important to assess residual confounding for non-zero and spatially structured trends. To do this we extended the residual confounding simulations to include trends that varied in direction, magnitude, and spatial patterns. Our strategy was to inform the simulation data generating process by using real data as much as possible while reducing the synthetic components (and assumptions therein) to a minimum (see Knaus et al. (2021) for other examples of empirically driven simulations.) The synthetic components in our simulation were based on two key assumptions, 1) that populations change at the same rate across the study period, and, 2) that spatial patterning aligned with edge-core population structure. Testing these assumptions is an area for further research.

**SECTION 4.1.2: Inferential Scope**
An important goal of this paper was to investigate the use of DML to control for interannual confounding when estimating trends based on citizen science data. The results show that DML can be used to reduce confounding bias leading to more accurate estimates and stronger associative inferences. These results are in line with other efforts that seek to improve associative inferences by harnessing approaches and ideas originally developed for causal inference (Bühlmann, 2020; Cui & Athey, 2022).

With additional assumptions DML can also be used for causal inference. The key assumption to make causal inference is to assert the absence of confounders that are missing from the analysis and independent of the original features (sometimes called *missing, hidden, or unmeasured confounders*). Practically, asserting the absence of missing confounders requires assumptions that go beyond the data in hand (Hernán & Robins, 2020). Neither the propensity score model nor the simulation-based residual confounding analysis can detect or control for missing confounders. Thus, end-users need to carefully consider the strength of their domain knowledge and the limits of inference.

**Section 4.2 Estimating Uncertainty**
Estimating uncertainty for the modelling of trends using eBird data presented two challenges that were not present in the synthetic example. The first challenge was to capture the sampling variation from both steps of the abundance hurdle model. To do this we used a resampling approach to propagate uncertainty from both estimation steps when estimating the 80% intervals. The second challenge was to assess the DML performance at estimating uncertainty based on temporally correlated counts. Temporal correlation in observed species count data can be induced by several important ecological processes, so it is important to include such correlations in the simulation study. This presents an analytical challenge the DML outcome model does not include structural components or features necessary to account for temporal correlation.

The simulation results showed that confidence interval coverage for the eBird analysis was below the nominal 80% level. We believe this is caused, at least partially, by the outcome model not accounting for the temporal correlation. Nevertheless, these same simulation results demonstrated that there was strong directional error control when we used the interval estimates to identify non-zero trends, i.e., trend estimates whose intervals did not contain zero. We interpret these two results to indicate that the uncertainty may not be scaling appropriately with the magnitude of the trend. Accounting for temporal correlation in the outcome is an interesting direction for further research into use of the DML framework. Residual spatial structure is another common feature of large-scale geographic studies that is absent from the DML. Given that the processes driving population change are more likely to vary locally when data come from large geographic extents (Rose et al., 2017), an important avenue of additional research is into accounting for nonstationarity of the drivers and confounders of trends. Incorporating more powerful rules to identify non-zero trends that control for multiple comparisons (e.g. false detection rate thresholding) and spatial correlation could also serve to improve the power of the approach and the scope of inference for species with weaker trend signals like Canada warbler.

**Section 4.3: Other Applications**
In this study we showed how spatial features can be used to estimate spatial patterns of variation in the trends (Fig.3, 4, SI-3, SI-4). The ability to associate features and trends can also be used to study other patterns of variation. For example, including indicators in the trend feature set can be used to estimate the effects of different survey protocols, management actions, or policies on population change. This could be useful for conservation planning, assessment, to inform integrated analysis, or for future survey design. Moreover, by including other features in the trend model that capture other potential trend effects (e.g. changes in landcover) it is possible to study systematic management differences after accounting for changes in landcover. This may be useful for Before-After-Control-Impact studies with citizen science data where accounting for the simultaneous impacts of other management and environmental changes is a challenge (Kerr et al., 2019). Finally, learned associations between trends and features can also be used to forecast expected population changes, conditional on a given set of features values and the assumption that the underlying processes driving population change during the study period will persist into the future.

**Section 4.4: Other Sources of Citizen Science Data**
The DML trend model and the simulation-based adjustment presented here can be used with other data types and applications. The Causal Forest implementation (Tibshirani et al., 2020) can accommodate binary and real valued outcome variables, making it possible to estimate trends in species occurrence rates, expected trends, and other indices of abundance.

For the eBird analysis presented here we used the fact that species counts were collected as part of a complete checklist of birds, which allowed us to infer the zero counts associated with non-detection. The same approach can be used with other checklist-based citizen science projects to analyse counts as well as binary response presence-absence data (e.g. birds BTO 2017; Swiss bird project; and butterflies van Swaay et al., 2008). Even when observations are not collected in the form of complete lists, for some taxa observations can be assembled into pseudo-checklists (Henckel et al., 2020; van Strien et al., 2013) making them amenable to DML trend analysis. It may also be possible to analyse presence-only data (e.g. iNaturalist.org) by carefully selecting (Valavi et al., 2021) or weighting (Fithian & Hastie, 2013) background data. However, more research will be needed to carefully consider biases and confounding associated with presence-only data (e.g. Stoudt et al., 2022). The DML trend model may even be useful for the analysis of data collected from structured surveys where confounding can arise despite survey structure (e.g. Zhang et al. 2021).

## SECTION 5: Conclusion

The volume of citizen science data is rapidly growing, but the lack of structured protocols has rendered most of these data unsuitable for estimating population trends. Bias can be introduced by changes over time in how people participate. The DLM trend model can account for these confounding changes over time. When used appropriately, including assessments of the propensity score model used to account for sources of confounding variation, this approach has the potential to increase the biodiversity monitoring value that we can obtain from citizen data. This could enable us to better track population changes in areas of the world with fewer structured monitoring programmes.

Imbens, G. W., & Rubin, D. B. (2015). *Causal inference in statistics, social, and biomedical sciences*. Cambridge University Press.
IUCN. (2019). *The IUCN Red List of Threatened Species. Version 2019.* https://www.iucnredlist.org
Johnston, A., Hochachka, W., Strimas-Mackey, M., Ruiz Gutierrez, V., Robinson, O., Miller, E., Auer, T., Kelling, S., & Fink, D. (2019). *Analytical guidelines to increase the value of citizen science data: Using eBird data to estimate species occurrence* [Preprint]. Ecology. https://doi.org/10.1101/574392
Johnston, A., Matechou, E., & Dennis, E. B. (2022). Outstanding challenges and future directions for biodiversity monitoring using citizen science data. *Methods in Ecology and Evolution*, *n/a*(n/a). https://doi.org/10.1111/2041-210X.13834
Kelling, S., Johnston, A., Bonn, A., Fink, D., Ruiz-Gutierrez, V., Bonney, R., Fernandez, M., Hochachka, W. M., Julliard, R., Kraemer, R., & Guralnick, R. (2019). Using Semistructured Surveys to Improve Citizen Science Data for Monitoring Biodiversity. *BioScience*, *69*(3), 170–179. https://doi.org/10.1093/biosci/biz010
Kerr, L. A., Kritzer, J. P., & Cadrin, S. X. (2019). Strengths and limitations of before–after–control–impact analysis for testing the effects of marine protected areas on managed populations. *ICES Journal of Marine Science*, *76*(4), 1039–1051. https://doi.org/10.1093/icesjms/fsz014
Kery, M., & Royle, J. A. (2020). *Applied Hierarchical Modeling in Ecology: Analysis of Distribution, Abundance and Species Richness in R and BUGS: Volume 2: Dynamic and Advanced Models*. Academic Press.
Knaus, M. C., Lechner, M., & Strittmatter, A. (2021). Machine learning estimation of heterogeneous causal effects: Empirical Monte Carlo evidence. *The Econometrics Journal*, *24*(1), 134–161. https://doi.org/10.1093/ectj/utaa014
Link, W. A., Sauer, J. R., & Niven, D. K. (2020). Model selection for the North American Breeding Bird Survey. *Ecological Applications*, *30*(6), e02137. https://doi.org/10.1002/eap.2137
Nie, X., & Wager, S. (2021). Quasi-oracle estimation of heterogeneous treatment effects. *Biometrika*, *108*(2), 299–319. https://doi.org/10.1093/biomet/asaa076
Obermeyer, Z., & Emanuel, E. J. (2016). Predicting the Future—Big Data, Machine Learning, and Clinical Medicine. *The New England Journal of Medicine*, *375*(13), 1216–1219. https://doi.org/10.1056/NEJMp1606181
Pocock, M. J. O., Tweddle, J. C., Savage, J., Robinson, L. D., & Roy, H. E. (2017). The diversity and evolution of ecological and environmental citizen science. *PLOS ONE*, *12*(4), e0172579. https://doi.org/10.1371/journal.pone.0172579
R Core Team. (2019). *R: A Language and Environment for Statistical Computing*. R Foundation for Statistical Computing. https://www.R-project.org/
Ramsey, D. S. L., Forsyth, David. M., Wright, E., McKay, M., & Westbrooke, I. (2019). Using propensity scores for causal inference in ecology: Options, considerations, and a case study. *Methods in Ecology and Evolution*, *10*(3), 320–331. https://doi.org/10.1111/2041-210X.13111
Robinson, P. M. (1988). Root-N-consistent semiparametric regression. *Econometrica: Journal of the Econometric Society*, 931–954.
Rose, K. C., Graves, R. A., Hansen, W. D., Harvey, B. J., Qiu, J., Wood, S. A., Ziter, C., & Turner, M. G. (2017). Historical foundations and future directions in macrosystems ecology. *Ecology Letters*, *20*(2), 147–157.
Rosenbaum, P. R., & Rubin, D. B. (1983). The central role of the propensity score in observational studies for causal effects. *Biometrika*, *70*(1), 41–55.
20

**Acknowledgments:** The eBird and the eBird Status and Trends projects rely on the time, and dedication and support from countless individuals and organizations. We thank the many thousands of eBird participants for their contributions and the eBird team for their support.

**Funding:** This work was funded by The Leon Levy Foundation, The Wolf Creek Foundation, and the National Science Foundation (ABI sustaining: DBI-1939187). This work used Bridges2 at Pittsburgh Supercomputing Center and Anvil at Rosen Center for Advanced Computing at Purdue University through allocation DEB200010 from the Advanced Cyberinfrastructure Coordination Ecosystem: Services & Support (ACCESS) program, which is supported by National Science Foundation grants #2138259, #2138286, #2138307, #2137603, and #2138296. Our research was also funded through the 2017-2018 Belmont Forum and BiodivERsA joint call for research proposals, under the BiodivScen ERA-Net COFUND program, with financial support from the Academy of Finland (AKA, Univ. Turku: 326327, Univ. Helsinki: 326338), the Swedish Research Council (Formas, SLU: 2018-02440, Lund Univ.: 2018-02441), the Research Council of Norway (Forskningsrådet, NINA: 295767) and the U.S. National Science Foundation (NSF, Cornell Univ.: ICER-1927646).

# Supplemental Information for
## A Double Machine Learning Trend Model for Citizen Science Data

Daniel Fink[1*], Alison Johnston[2], Matt Strimas-Mackey[1], Tom Auer[1], Wesley M. Hochachka[1], Shawn Ligocki[1], Lauren Oldham Jaromczyk[1], Orin Robinson[1], Chris Wood[1], Steve Kelling[1], and Amanda D. Rodewald[1]

[1]Cornell Lab of Ornithology, Cornell University, Ithaca, NY 14850, USA.
[2]Centre for Research into Ecological and Environmental Modelling, School of Maths and Statistics, University of St Andrews, St Andrews, UK.
*Corresponding author. daniel.fink@cornell.edu

**Contents**
This supplemental Information document contains the following sections:
1) DML Literature Review
2) The Potential Outcomes Framework
3) Data Description
4) Abundance Ensemble Model
5) eBird Simulation
    a. Overview
    b. Generative Model
    c. Simulation Scenarios
    d. The Discrete-time stochastic exponential growth rate model
6) Directional Error
7) Simulation figures for Canada warbler and long-billed curlew
8) References for Supplemental Information

**Section S1: DML Literature Review**
Causal machine learning is a broad, active field of research (Kaddour et al., 2022) where machine learning methods are employed to reason about the factors that affect data generating processes (interventions or exposures) and what would have happened in hindsight (counterfactuals). This includes work adapting standard machine learning methods to flexibly estimate heterogenous treatment effects along a potentially large number of covariates. Recent advances for estimating heterogenous treatment effects include methods based on random forests (Athey et al., 2019; Wager & Athey, 2018) the LASSO (Chen et al., 2017), Bayesian additive regression trees (Hahn et al., 2020), boosting (Powers et al., 2018), neural networks (Shalit et al., 2017), and metalearners based on combinations of these methods (Künzel et al., 2019). DLM approaches include (Chen et al., 2017; Chernozhukov et al., 2018b; Colangelo & Lee, 2022; Foster & Syrgkanis, 2020; Jung et al., 2021; Kennedy, 2022; Nie & Wager, 2021). Carvalho et al. (2019) and Knaus et al.(2021) have compared the empirical performance of several of these methods. Several packages are available including the grf R package for Causal Forests (Tibshirani et al., 2020) the CausalML package for DML is available in R and Python (Chen et al., 2020) and the EconML package is available in Python (Battocchi, 2019).

**Section S2: The Potential Outcomes Framework**

The potential outcomes framework (Rubin, 1974) provides a theoretical basis for causal inference using counterfactuals (outcomes that did not occur but would likely have occurred if the cause had occurred differently) when analyzing the relationship between cause and effect under different exposure or treatment conditions. The outcome model in Equation 1 (and equation S1) is closely connected to the potential outcomes framework where we posit potential outcomes $\big(Y_i(T=1), Y_i(T=2), \ldots, Y_i(T=p)\big)$ corresponding to the species abundance $Y_i$ on the $i^{th}$ survey that would have been reported had it been conducted on different years of the study period, $T = (1, \ldots, p)$. In this setup, the average per year rate of change in population size is estimated as the exposure or treatment effect.

Under the following assumptions the trend, $\tau$, can be estimated without bias in the potential outcomes framework (Ramsey et al., 2019) . First, potential outcomes are independent of the observation year conditional on features $X$, the set of potential confounders. This assumption has two important implications. The first implication is that there are no hidden or missing confounders. For this reason, this assumption is often termed the *no unmeasured confounding*, *conditional ignorability*, or *unconfoundedness* assumption (Rosenbaum & Rubin, 1983). Ultimately, justifying this assumption requires subject-matter knowledge that extends beyond the data in hand and can be particularly challenging to justify in applications where subject-matter knowledge may be insufficient to identify all important confounders (Hernán & Robins, 2020).

The second implication of the unconfoundedness assumption is that the propensity score model adequately captures the interannual variation necessary to achieve the conditional independence of the observation year given features $X$. Unlike the first implication of the unconfoundedness assumption, this is a modeling assumption that can be checked with data and is the motivation the residual confounding diagnostic and adjustment proposed in Section 2.2 of the main paper.

The second assumption states that each survey must have nonzero probability of being surveyed in any year during the study period $T = (1, \ldots, p)$. This follows from the logic that if a given survey could only be conducted on a single year, then the trend, defined as a difference of potential outcomes, would be undefined. Practically, this assumption can be checked by investigating the degree of overlap or common support in the distribution of features among years in the propensity score model (Ramsey 2019). This assumption is often termed the *common support* assumption.

The third assumption states that the outcome reported for a given survey is not affected by the year of observation assigned to other checklists. This assumption is known as the *stable unit treatment value assumption* (Rubin, 1980) and generally requires that treatments or exposures at one unit do not affect the outcomes for other units.

**Section S3: Data description**
In this supplement we provide a detailed description of the data and data processing, including the aggregation, used for the eBird analysis.

We estimated species trends using data from eBird, a popular citizen science project that has been collecting bird observation data since 2002 (Sullivan et al., 2014). eBird is a semi-structured survey (Kelling et al., 2019) because its flexibility allows participants to collect data in the ways they choose, but auxillary data are collected that describe the variation in data collection methods. There are two important components of data collection that are followed by many eBird participants. First, participants are encouraged to report all bird species detected and identified during the survey period, resulting in a *complete checklist* of bird species. This limits variation in preferential reporting rates across species and provides a basis to infer species non-detections. Second, participants are also encouraged to report characteristics of their birdwatching, for example the time of day and distance travelled. These variables and others can be used to adjust for variation in detection rates (Johnston et al., 2019).

The bird observation data were obtained from the citizen science project, eBird (Sullivan et al., 2014). We used a subset of data in which the time, date, and location of each survey were reported, and observers recorded the number of individuals of all bird species detected and identified during the survey period, resulting in a *complete checklist* of species on the survey (Sullivan et al., 2009). Only the first participant's checklist was considered for each group of linked ("shared") checklists; shared checklists are produced by the duplication of an original checklist followed by reassigning the copy to a different observer who was part of the same group of observers during the period of observation. We further restricted checklists to those collected with 'stationary' or 'traveling' protocols from January 1, 2007 to December 31, 2021. Traveling surveys were restricted to those ≤ 10km. The resultant dataset consisted of 43,814,030 checklists. Each species' trend analysis and simulations were based on breeding season subsets of this dataset. We used the species-specific breeding season range and dates published in (Fink, Auer, Johnston, Strimas-Mackey, et al., 2020).

The observation effort variables were: (a) the duration spent searching for birds, (b) whether the observer was stationary or traveling, (c) the distance traveled during the search, (d) the number of people in the search party, and (e) the Checklist Calibration Index (CCI), a standardized measure indexing differences in the rate at which observers accumulate new species, controlling for geographic region, habitat and other factors that determine the number of species that potentially could be observed (Kelling et al., 2015). Note, only the first observer's CCI was associated with checklists that contain more than one observer.

The temporal variables included observation time of the day, which was standardized across time zones and daylight savings times as the difference from solar noon, was used to model variation in availability for detection, e.g. variation in behavior such as participation in the dawn chorus (Diefenbach et al., 2007). The day of the year (1-366) on which the search was conducted was used to capture intra-annual variation and the year of the observation was included to account for inter-annual variation.

Spatial and spatiotemporal descriptors of the local environment were variables describing elevation, topography, shorelines, islands, land cover, land use, hydrology, and road density (Meijer et al., 2018). To account for the effects of elevation and topography, each checklist location was associated with elevation (Tozer et al., 2019), eastness, and northness. These latter two topographic variables combine slope and aspect to provide a continuous measure describing geographic orientation in combination with slope at both $90m^2$ and $1km^2$ resolutions (Amatulli et al., 2018). Each checklist was also linked to a series of covariates derived from the NASA MODIS land cover, land use, and hydrology data; MCD12Q1 (Friedl & Sulla-Menashe, 2019). We selected this data product for its moderately high spatial resolution and annual temporal resolution to capture spatial patterns of change in land cover, land use, and hydrology. We used the FAO-Land Cover Classification System which classifies each 500m pixel into land cover one of 21 vegetative cover classes, along with additional classifications describing the land use and hydrology of each pixel. Checklists were linked to the MODIS data by year from 2001-2020, capturing inter-annual changes in land cover. The checklist data for 2021 were matched to the 2020 data, as MODIS data from after 2020 were unavailable at the time of analysis. Additionally, to delineate the interface between terrestrial, aquatic, and marine environments we used NASA MODIS land water classification MOD44W (Carroll et al., 2017, p. 44) in conjunction with 30m shoreline and island data (Sayre et al., 2019) and the elevation data described above to classify each pixel into land, ocean, inland water, and coastal areas. To identify habitat for coastal species, tidal mudflats were classified in three-year windows (Murray et al., 2019). Finally, hourly weather variables were assigned from the Copernicus ERA5 reanalysis product at 30km resolution (Hersbach et al., 2020).

To prepare these data for the trend analysis we aggregated all checklists, separately for each year, using a spatio-temporal grid whose dimensions were (27km × 27km x 1week) based on the latitude, longitude, date, and year of each checklist. The response variable for each grid cell was the mean count of the given species averaged for all checklists within each grid cell. Grid cells without any detections of this species during any weeks during any years were removed. Averaging counts in this way strengthens abundance signals making it easier to detect trends among species that are detected less frequently and species with relatively low abundance.

The feature sets were based on the variables in Table **SI-1.** We aggregated all observation effort variables by summing to represent the total effort within each grid cell. We also calculated the mean search duration and distance within each grid cell. Grid cell CCI values were computed as the average CCI within each grid cell weighted by search duration. The grid cell mean values were calculated for the time of day when surveys were initiated and for the day of the year when surveys were conducted. To keep track of the survey year, we also created a new variable for each grid cell indicating the unique year when surveys were conducted within the cell.

All spatial and spatiotemporal variables were summarized within grid cells using two metrics. The first metric describes the mean or *composition* of the variable across the grid cell landscape. The second metric describes the variation or *spatial configuration* of the variable across the grid cell landscape. For the categorical class variables, we computed the composition as the proportion of each class within each grid cell (PLAND) and we computed the spatial configuration using an index of its edge density within each grid cell (ED) using the R package *landscapemetrics* (Hesselbarth et al., 2019; McGarigal et al., 2012) . For the continuous variables (elevation, eastness, and northness) we computed the median and standard deviations for the composition and configuration, respectively.

| Data Description | Data Source |
|---|---|
| Longitude | Sullivan et al. 2014 |
| Latitude | Sullivan et al. 2014 |
| Year | Sullivan et al. 2014 |
| Day | Sullivan et al. 2014 |
| Solar Noon Time | Sullivan et al. 2014 |
| Search Effort Duration (hrs) | Sullivan et al. 2014 |
| Search Effort Distance (km) | Sullivan et al. 2014 |
| Number of Observers | Sullivan et al. 2014 |
| CCI | Sullivan et al. 2014 |
| Elevation (m) | Becker et al. 2009 |
| Eastness slope/aspect component | Becker et al. 2009, Amatulli et al. 2018 |
| Northness slope/aspect component | Becker et al. 2009, Amatulli et al. 2018 |
| Island indicator | Sayre et al. (2018) (30) |
| Tidal Mudflats | Murray et al. (2019) (31) |
| Ocean | Carroll et al. 2017, Sayre et al. (2018) |
| River | Carroll et al. 2017, Sayre et al. (2018) |
| Lakes | Carroll et al. 2017, Sayre et al. (2018) |
| Barren | Friedl & Sulla-Menashe 2019 |
| Permanent Snow and Ice | Friedl & Sulla-Menashe 2019 |
| Evergreen Needleleaf Forests | Friedl & Sulla-Menashe 2019 |
| Deciduous Broadleaf Forests | Friedl & Sulla-Menashe 2019 |
| Mixed Broadleaf/Needleleaf Forests | Friedl & Sulla-Menashe 2019 |
| Mixed Broadleaf Evergreen/Deciduous Forests | Friedl & Sulla-Menashe 2019 |
| Open Forests | Friedl & Sulla-Menashe 2019 |
| Sparse Forests | Friedl & Sulla-Menashe 2019 |
| Dense Herbaceous | Friedl & Sulla-Menashe 2019 |
| Sparse Herbaceous | Friedl & Sulla-Menashe 2019 |
| Dense Shrublands | Friedl & Sulla-Menashe 2019 |
| Shrubland/Grassland Mosaics | Friedl & Sulla-Menashe 2019 |
| Sparse Shrublands | Friedl & Sulla-Menashe 2019 |
| Forest/Cropland Mosaics | Friedl & Sulla-Menashe 2019 |
| Natural Herbaceous/Croplands Mosaics | Friedl & Sulla-Menashe 2019 |
| Herbaceous Croplands | Friedl & Sulla-Menashe 2019 |
| Woody Wetlands | Friedl & Sulla-Menashe 2019 |
| Herbaceous Wetlands | Friedl & Sulla-Menashe 2019 |
| Tundra | Friedl & Sulla-Menashe 2019 |
| Urban and Built-up Lands | Friedl & Sulla-Menashe 2019 |
| Highways | Meijer et al. 2018 |
| Primary Roads | Meijer et al. 2018 |
| Secondary Roads | Meijer et al. 2018 |
| Tertiary Roads | Meijer et al. 2018 |
| Local Roads | Meijer et al. 2018 |
| Wind Speed (E-W) m/s | Hersbach et al. 2021 |
| Wind Speed (N-S) m/s | Hersbach et al. 2021 |
| 2m Dewpoint (C) | Hersbach et al. 2021 |
| 2m Temperature (C) | Hersbach et al. 2021 |
| High Cloud Cover (%) | Hersbach et al. 2021 |
| Medium Cloud Cover (%) | Hersbach et al. 2021 |
| Low Cloud Cover (%) | Hersbach et al. 2021 |
| Instantaneous Wind Gust (m/s) | Hersbach et al. 2021 |
| Snowfall, water equivalent (m) | Hersbach et al. 2021 |
| Rainfall (m) | Hersbach et al. 2021 |
| Sea Level Pressure, mean (Pa) | Hersbach et al. 2021 |
| Sea Level Pressure, change (Pa) | Hersbach et al. 2021 |

**Table SI-1** Variables used in trend analysis.

**SECTION S4: The Abundance Ensemble**
The ensemble was generated by refitting the causal forests with training datasets randomized to capture sampling variation and variation arising from the randomized spatiotemporal aggregation procedure. First, 75% of the available checklists we randomly sampled. Second, we aggregate the sampled checklists using a randomly located grid. To generate additional independence between the occurrence and count stages of the hurdle model, the training data were independently sampled and aggregated for the occurrence and count model training. We computed 10 ensemble estimates each for the occurrence and count models. To estimate trends in population abundance and their uncertainty we considered the ensemble consisting of all 100 unique combinations of occurrence and count estimates. We did not choose to compute more ensemble members because of the high computational cost of computing ensemble estimates for the simulation analyses.

**Section S5: eBird Simulation**
In this supplement we provide additional details about eBird simulation used for the residual confounding analysis and model assessment. This section includes an overview of the simulation construction and subsections on the generative model, the simulation scenarios, and the discrete-time stochastic growth rate model.

**Section S5.1 Simulation Overview**
These simulations were constructed to meet four objectives: 1) To generate realistic patterns of each species' occurrence and counts on eBird checklists; 2) with specified trends in abundance; 3) including temporal correlation that arises from environmental stochasticity in population growth rates; and 4) replicating the interannual confounding in the eBird data.

The following steps were used to generate the simulated datasets for each species (Figure S1-1):

1) **Train generative model to learn species' occurrence and abundance**. We trained a predictive model (Fink, Auer, Johnston, Ruiz-Gutierrez, et al., 2020) to learn and then generate realistic patterns of variation in species' occurrence and counts on eBird checklists that were constant across years. To ensure that the generative model learned patterns constant across years it was trained using a random subset of the species' data with a permuted version of the observation year feature: $Year_p$. This model learned patterns for both the eBird observation process (e.g., variation in detection rates associated with varying amounts of search effort) and species ecological processes (e.g., variation in occurrence and counts associated with environmental features).
2) **Generate simulated checklists, $X^*$.** To replicate the interannual variation in the original features $X$ we generated the simulated checklists $X^*$ by resampling the original features $X$ with replacement, stratified by year. The stratification by year ensures that any interannual changes in sampling are reflected in the simulated checklists $X^*$.
3) **Simulate species trends**. Using checklists $X^*$ we initialized the simulated population for the first year of the study using the expected species occurrence rates and counts predicted from the generative model. Then population dynamics were simulated based on the permuted observation year feature, $Year_p$, using a discrete-time stochastic exponential growth model with a spatially explicit growth rate. Ten different simulation scenarios were created for each species with different patterns of population increase and decrease. The expected growth rate at each location was constant across years and varied according to the given simulation scenario. The ten simulation scenarios were generated, varying in direction, magnitude, and spatial patterns. Magnitudes were set to

<1% (weak), 3.3% (moderate) and 6.7% per year (strong) based on IUCN Red List criteria (IUCN, 2019). The spatially varying trends were constructed to vary in direction and magnitude along a gradient from the core to the edge of the species' population (See Figure 4 and Figures SI-2 and SI-3). Finally, we included a stochastic component of the simulated growth rate to replicate the temporal correlation expected from interannual environmental stochasticity, an important characteristic of real species abundance data. The stochastic component within each cell was drawn from a random distribution with the same cell-specific mean, with distributions varying independently among 27km grid cells and years with a standard deviation of approximately 6% per year.

4) **Simulate observations.** The observed checklist counts were simulated from the trajectories of expected population abundance created in step 3. The reported species detection was generated as a realization from a Bernoulli distribution based on the expected occurrence rate. Conditional on detection, we generated the reported checklist count as a realization from a Poisson distribution with rate set to be the logarithm of the expected count at each site in each year.

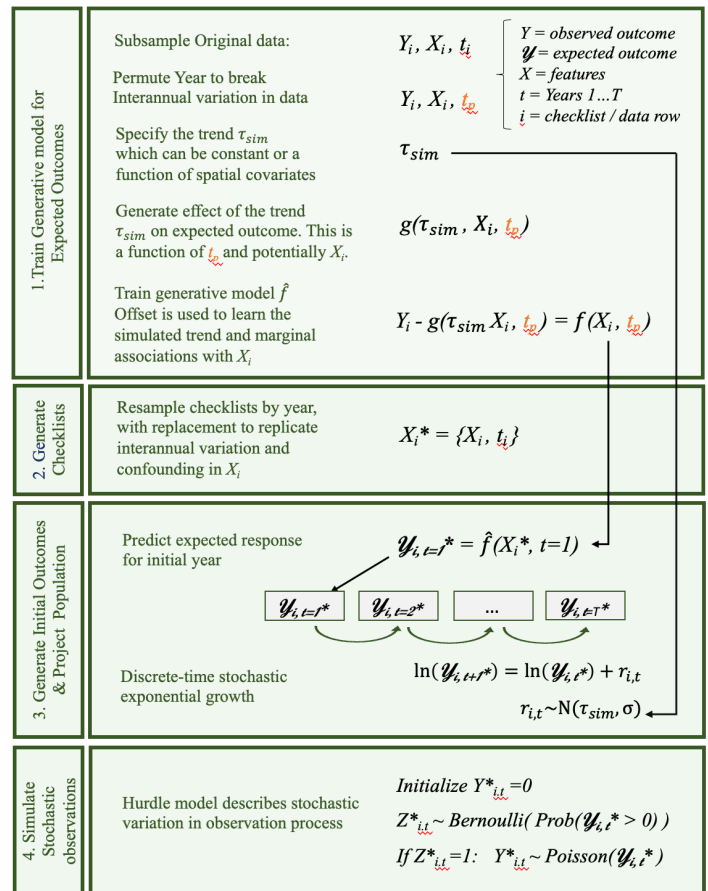

**Figure S1-1: A schematic illustration of the data generating process used to simulate species-specific data with specified trends and data-driven confounding.** Each species simulation begins with a subset of data and progresses through the four steps described in Section SI-5. The data generating process was informed as much as possible by real data to capture the full range of confounding while reducing the scope of simplifying assumptions necessary to generate synthetic trends.

**Section S5.2: The Generative model**
The goal of the generative model was to train a model to predict expected patterns of species abundance based on user-specified population trends along with realistic ecological and observational patterns learned from data. To do this we used the Boosted Regression Trees (BRT) hurdle model described in Fink, Auer, Johnston, Ruiz-Gutierrez, et al., (2020).

Let $(N, Y, X_e, X_o, year)$ be the set of training data for a given region, season, and species where:
- $N$ is the $n \times 1$ vector of observed counts on the *n* surveys in the training data,
- $Y$ is the $n \times 1$ vector that indicates the checklists with count greater than zero,
- $X_e$ is the $n \times k$ matrix of *k* predictors that describe the ecological process,
- $X_o$ is the $n \times j$ matrix of *j* predictors that describe the observation process, and
- $year$ is the $n \times 1$ vector of the year each survey was conducted.

We begin with the BRT hurdle model for species abundance and then explain how it was modified for generative modeling. In the first step of the unmodified BRT hurdle model, a Bernoulli response BRT is trained to predict the probability of occurrence:
$$Y \sim Bernoulli(\pi)$$
$$logit(\pi) = f(X_e, X_o, year)$$
where $\pi$ is the probability of occurrence and the function $f()$ is fit using boosted decision trees. In the second step, the Poisson response BRT,
$$N \sim Poisson(\mu)$$
$$log(\mu) = f(X_e, X_o, year)$$
is trained to predict the expected counts $\mu$, using the subset of the training data observed to be present, i.e. $N > 0$.

To generate simulated counts of reported birds we modify the BRT as follows. The first modification permutes the $year$ predictor variable. This ensures that the BRTs cannot learn year-to-year variation from the training data. The second modification trains the hurdle BRT with an offset constructed to impart a user-specified trend when training the Poisson response model. The modified fitting procedure begins with the Bernoulli response BRT, trained with the permuted year,
$$Y \sim Bernoulli(\pi)$$
$$logit(\pi) = f(X_e, X_o, year^p),$$
The Poisson response BRT is trained with the permuted year and the offset, $O$,
$$N \sim Poisson(\mu)$$
$$log(\mu) + O = f(X_e, X_o, year^p).$$
The offset can be considered as an adjustment to the expected observed counts, on the log-link scale. We construct the offset $O = g(year^p)$ where $g()$ is a function of the permuted year value, $year^p$. Because the offset is the only source of interannual variation with respect to $year^p$, it forces the boosting procedure to learn the user specified population trend.

**Section S5.3: The Simulation Scenarios**
The three types of spatial trend offsets constructed were: 1) spatially constant trends, 2) spatially varying trends and 3) no trend. We used the following linear model to construct the trend offsets, $O = \alpha\, year^p + \alpha_I year^p X_I$, where $\alpha$ controls the strength and direction of the overall year-to-year changes in the expected log count and $\alpha_I$ controls the strength of the interaction between $year$ and $X_I$, the interacting variable. To avoid scaling issues, we assume that the interacting variable has been transformed to vary between -1 and 1. Note that because an intercept is fit as part of $f()$, we do not include an intercept term in the offset.

Spatially uniform trends were generated by setting $\alpha_I = 0$. Trends that affect a population uniformly over a region may indicate the indirect effects of broad-spatial scale processes like climate change. Spatially varying trends can be generated by setting $\alpha = 0$ and specifying a spatially patterned variable $X_I$ to interact with $year^p$. To assess if spatial patterns associated with density dependent population processes can be detected, the spatially interacting variable was selected to be the land cover composition feature with highest importance score from the occurrence model and that had sufficient detections of the focal species across a range of land cover values. In this way the interacting variable functions as an index of population density. Processes like habitat loss, disease, and dispersal can interact with population density to generate spatially varying trend patterns, e.g. (Massimino et al., 2015).

The spatially constant trend values were generated by using annual changes in the count part of the model on the log scale defined as values $\alpha$ =(-0.08, -0.04, 0.04, 0.08). The largest absolute values for the spatially constant trend (-0.08 and 0.08) were selected to generate spatially constant trends of approximately 6.7% per year averaged across species' ranges. With the spatially variable trends, positive and negative directions generated inversely related trend patterns, but both varying in association with the same landcover covariate (See Figure 2 in the main text and Figures SI-2 and SI-3). The spatially varying trend values were generated with the parameter interacting with landcover varying across values $\alpha_I$ = (-0.40, -0.20, 0.20, 0.40), with the largest values selected to generate relatively large regions within the species' range experiencing changes in population size of at least 6.7% per year.

### SECTION S5.4: The Discrete-time stochastic exponential growth rate model

In this supplemental section we describe the discrete-time stochastic exponential growth rate model used to incorporate population dynamics and temporal correlation into the simulated data sets. To set notation, we begin with the deterministic discrete-time exponential growth model,
$$N_{t+1,s} = (1 + r_s)N_{t,s}$$
where
- $N_{t,s}$ is an index of population size at time *t* and location *s*. In this study, $N_{t,s}$ is the count of the species on the checklist indexed by year *t* and location *s*,
- $r_s$ is the growth rate at location s. In our simulations $r_s$ is constant across years. We describe the rate of population change in terms of the *Percent Per Year* ($PPY$) change, so that $r_s = PPY/100$, and
- *t* indexes years *1…T* of the study period.

The stochastic growth rate model can be written as
$$N_{t+1,s} = R_{t,s}N_{t,s},$$
where $R_{t,s}$ is a random variable with $E(R_{t,s}) = (1 + r_s)$, the deterministic component of population growth, and $sd(R_{t,s})$, that captures the stochastic component of population growth.

To parameterize the model, it is convenient to take logarithms,
$$lnN_{t+1,s} = ln\,N_{t,s} + lnR_{t,s}.$$
We let $lnR_{t,s} \sim N\left(\mu_{R,s}, \sigma_R^2(s)\right)$ and write
$$lnN_{t+1,s} = ln\,N_{t,s} + \mu_{R,s} + \epsilon_t(s),$$
Where $\mu_{R,s'} = E[lnR_{t,s}]$ and $\epsilon_t(s) \sim N\left(0, \sigma_R^2(s)\right)$. Assuming that locations are indexed according to a spatial grid, the stochastic component of $R_{t,s}$ varies independently among spatial grid cells and among years. This process can represent environmental stochasticity like the year-to-year effects of extreme weather on reproductive success. Populations within the same grid cell experience the same environmental conditions affecting population growth. Despite the year-to-year independence of the stochastic effects, the population trajectories generated by this process will be temporally correlated within each grid cell due to the compounded effects of the stochasticity across years.

To generate realizations from the discrete-time stochastic exponential growth rate model two parameters need to be specified: $r_s$ describing the deterministic component of population growth and $\sigma_R(s)$ describing the stochastic component of population growth. For each simulation scenario, the values of $r_s$ are constructed from the generative model. To specify $\sigma_R(s)$ consider the following relationship between parameters of the normal and lognormal distributions,

$$sd(R_{t,s}) = E(R_{t,s})\sqrt{(\exp(\sigma_R^2(s)) - 1)}.$$

Because $r_s = \frac{PPY}{100}$, the range of values in which we are interested is relatively small, i.e. closer to 0.0 than to 1.0 in magnitude. For example, consider the 6.7 and 3.2 percent per year values used as IUCN RedList criteria (IUCN, 2019) and in the simulations. It follows that the standard deviation of $r_s$ will likely be small, i.e. closer to 0.0 than to 1.0. Using the assumption that these values should be small, and that $R_{t,s} = 1 + r_s$, we know that $E(R_{t,s}) \approx 1$, $\sqrt{(\exp(\sigma_R^2(s)) - 1)} \approx \sigma_R(s)$ and that $sd(R_{t,s}) = sd(r_s)$. Using these facts and substituting into the expression for $sd(R_{t,s})$ yields $sd(R_{t,s}) = \sigma_R(s)$. Thus, $100\sigma_R(s)$ is approximately the standard deviation in units of percent-per-year population growth.

For all simulation scenarios we specified $100\sigma_R(s)$ to be six percent per year. We chose this value to achieve a coefficient of variation for the growth rate of approximately 1 for the strong magnitude simulation scenarios and a coefficient of variation less than one for moderate and weak scenarios.

**SECTION S6 Directional Error**
In this supplemental section we describe how we assessed estimates of the trend direction (increasing or decreasing). Directional errors were defined to occur when non-zero trends estimated the direction incorrectly. Non-zero trends were defined to occur when the 80% confidence interval did not contain zero. Filtering out non-zero locations was an important step because estimating trends with sparse, noisy data is a difficult estimation task. This can be seen from the size of the uncertainty estimates (Figure 3 in the main text).

Like many classification tasks, we expected that the ability to detect non-zero trends (a measure of power) and the ability to avoid directional errors (false positive errors) would vary with the magnitude of the trend, reflecting the underlying signal-to-noise ratio. Understanding the associations between power and error rates and trend magnitude is useful for interpreting simulation performance and generalizing to the analysis of real data. To do this we summarized the proportion of interval estimates overlapping zero and the proportion of directional errors as a function of estimated trend magnitude, binned into half percent increments. This summarization was carried out across all 27km locations for 10 realizations from each of the 5 simulation scenarios held aside for model assessment.

Figure SI-1 shows how the power to detect non-zero trends increases rapidly with trend magnitude and the directional error rate decreases with trend magnitude across all simulations for all three species.

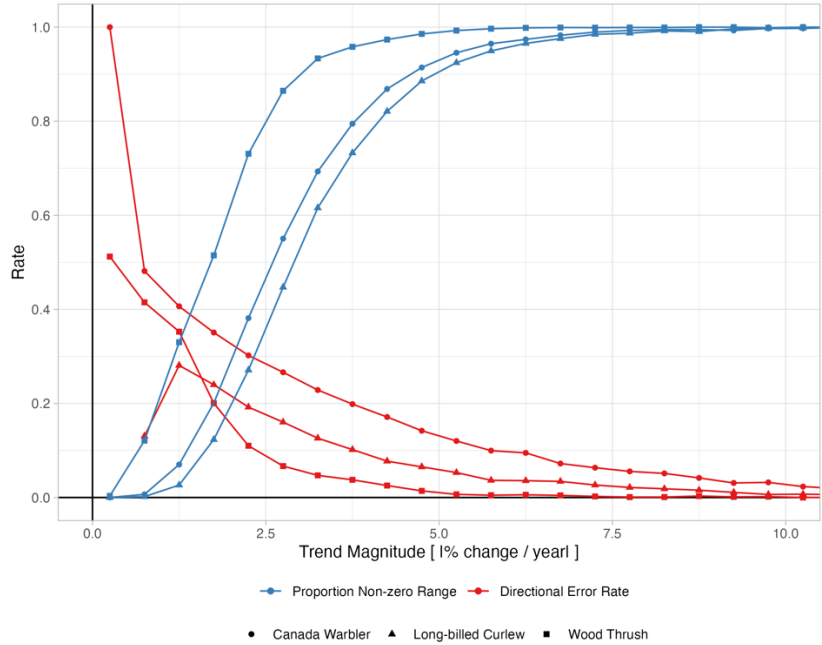

**Figure SI-2 Power detecting non-zero trends and directional error rates as functions of trend magnitude.** Blue lines show the power to detect non-zero trends as a function of trend magnitude binned into half Percent Per Year (PPY) increments for wood thrush (squares), Canada warbler (triangles), and long-billed curlew (circles). Red lines show the directional error rate as a function of trend magnitude binned into half Percent Per Year (PPY).

To summarize the species-level directional error rate across simulations we chose to report the mean directional error rate, averaged by magnitude binned into half-degree increments (Table 2 in main text). Implicitly, this measure gives each magnitude bin equal importance. This is a useful property when comparing analyses because it avoids confounding differences in the distribution of simulated trend magnitudes with the difference in error rates we are interested in.

## SECTION SI-7 Simulation Results for Canada warbler and long-billed curlew

This section includes figures showing trend maps for each of the ten scenarios for Canada warbler and long-billed curlew.

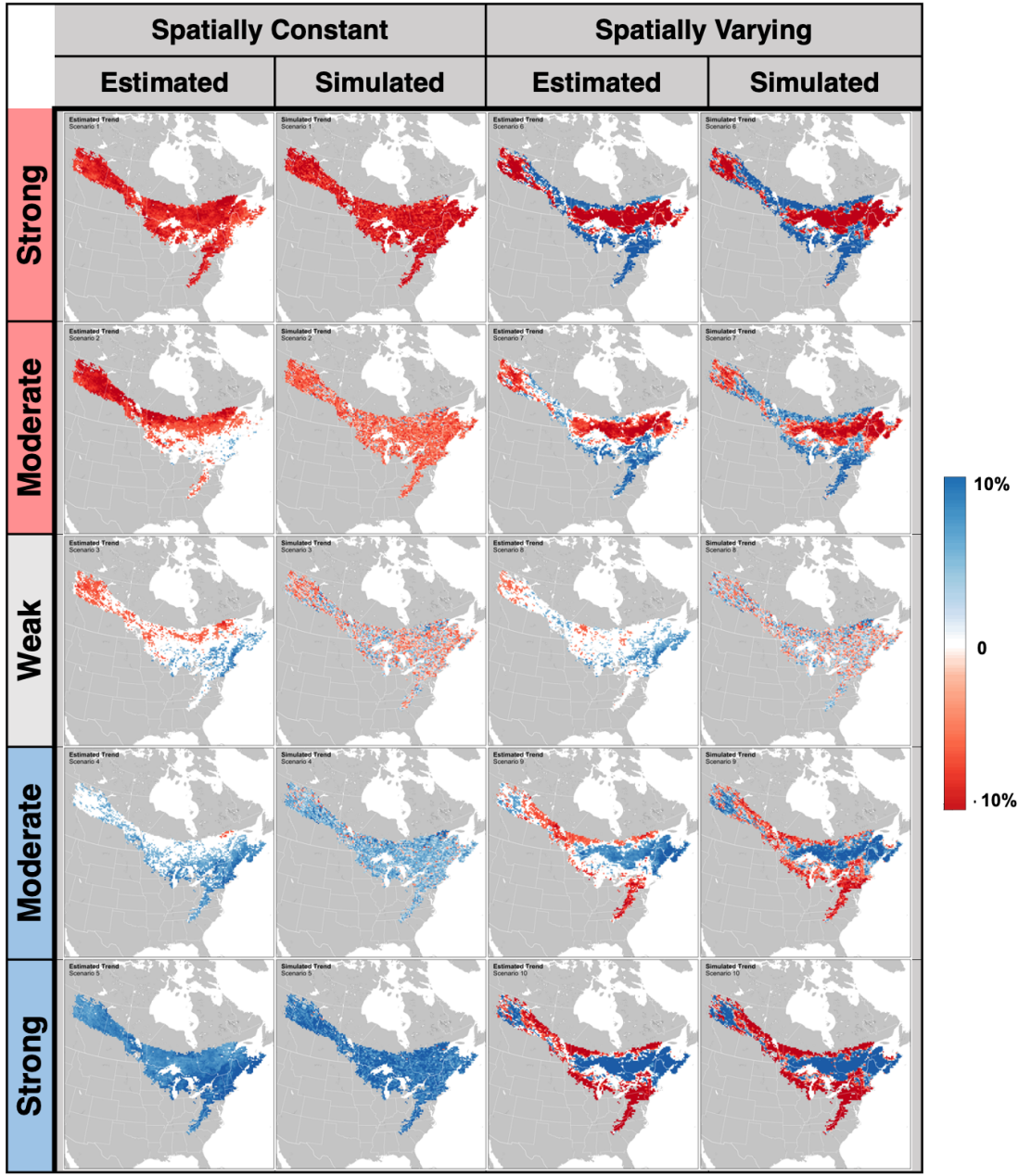

**Figure SI-3: Canada Warbler Trend Simulations.** All trend maps show the average annual percent-per-year change in abundance from 2007–2021 within 27km pixels (red=decline, blue=increase, white=non-significantly different from 0 at alpha =0.2), intensity (darker colors indicate stronger trends). Simulated trends show scenarios varying by direction and magnitude along rows: weak (includes trends $\sim|1\%/yr|$), moderate (includes regions with trends $\sim|3.5\%/yr|$), and strong trends (includes regions with trends $\sim|6.7\%/yr|$). The columns show simulated and estimated trends for spatially constant and varying simulation scenarios.

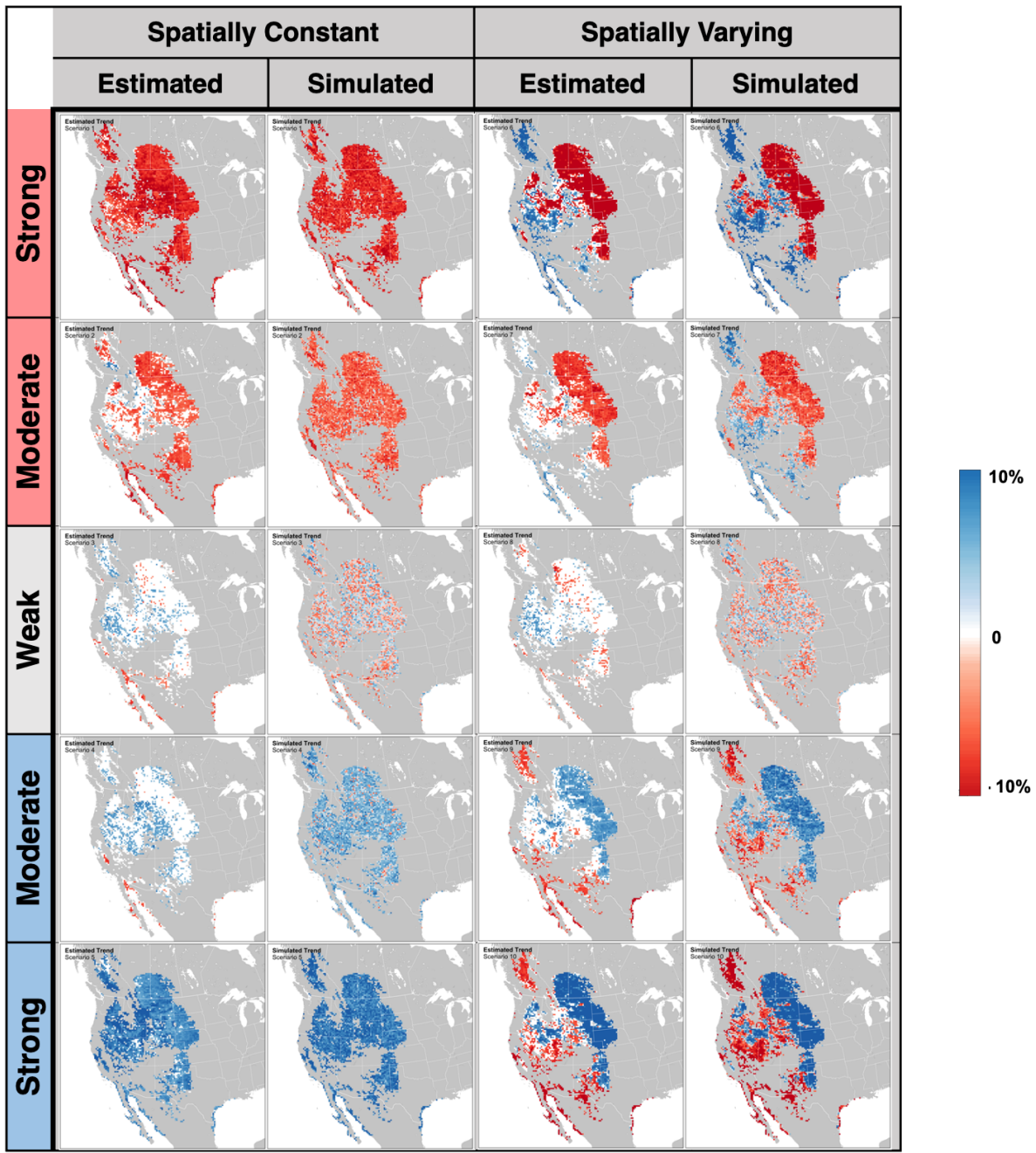

**Figure SI-4: Long-Billed Curlew Trend Simulations.** All trend maps show the average annual percent-per-year change in abundance from 2007–2021 within 27km pixels (red=decline, blue=increase, white=non-significantly different from 0 at alpha =0.2), intensity (darker colors indicate stronger trends). Simulated trends show scenarios varying by direction and magnitude along rows: weak (includes trends ~$|1\%/yr|$), moderate (includes regions with trends ~$|3.5\%/yr|$), and strong trends (includes regions with trends ~$|6.7\%/yr|$). The columns show simulated and estimated trends for spatially constant and varying simulation scenarios.

**Section SI 8: References for Supplemental Information**

Rubin, D. B. (1980). Randomization Analysis of Experimental Data: The Fisher Randomization Test Comment. *Journal of the American Statistical Association*, *75*(371), 591–593. https://doi.org/10.2307/2287653

Sayre, R., Noble, S., Hamann, S., Smith, R., Wright, D., Breyer, S., Butler, K., Van Graafeiland, K., Frye, C., & Karagulle, D. (2019). A new 30 meter resolution global shoreline vector and associated global islands database for the development of standardized ecological coastal units. *Journal of Operational Oceanography*, *12*(sup2), S47–S56.

Shalit, U., Johansson, F. D., & Sontag, D. (2017). Estimating individual treatment effect: Generalization bounds and algorithms. *Proceedings of the 34th International Conference on Machine Learning*, 3076–3085. https://proceedings.mlr.press/v70/shalit17a.html

Sullivan, B. L., Aycrigg, J. L., Barry, J. H., Bonney, R. E., Bruns, N., Cooper, C. B., Damoulas, T., Dhondt, A. A., Dietterich, T., & Farnsworth, A. (2014). The eBird enterprise: An integrated approach to development and application of citizen science. *Biological Conservation*, *169*, 31–40.

Sullivan, B. L., Wood, C. L., Iliff, M. J., Bonney, R. E., Fink, D., & Kelling, S. (2009). eBird: A citizen-based bird observation network in the biological sciences. *Biological Conservation*, *142*(10), 2282–2292.

Tibshirani, J., Athey, S., & Wager, S. (2020). *Grf: Generalized Random Forests. R package version 120*.

Tozer, B., Sandwell, D. T., Smith, W. H. F., Olson, C., Beale, J. R., & Wessel, P. (2019). Global Bathymetry and Topography at 15 Arc Sec: SRTM15+. *Earth and Space Science*, *6*(10), 1847–1864. https://doi.org/10.1029/2019EA000658

Wager, S., & Athey, S. (2018). Estimation and inference of heterogeneous treatment effects using random forests. *Journal of the American Statistical Association*, *113*(523), 1228–1242.
37